\documentstyle[12pt,subeqn,cite]{article}
\input epsf
\begin{document}
\baselineskip=21pt
%
\newcommand{\bc}{\begin{center}}
\newcommand{\ec}{\end{center}}
\newcommand{\be}{\begin{equation}}
\newcommand{\ee}{\end{equation}}
\newcommand{\bq}{\begin{eqnarray}}
\newcommand{\eq}{\end{eqnarray}}
\newcommand{\bsq}{\begin{subequations}}
\newcommand{\esq}{\end{subequations}}
\begin{titlepage}
\rightline {DTP/00/85}
\rightline {gr-qc/0009102}
\bigskip
\bc
{\Large \bf Cosmic strings in axionic-dilatonic gravity.}
\ec
\vskip0.5cm
\bc
Caroline Santos

Departamento de F\'\i sica da Faculdade de Ci\^encias da Universidade do
Porto,
Rua do Campo Alegre 687, 4150-Porto, Portugal

Centre for Particle Theory, Department of Mathematical Sciences,
South Road, Durham, DH1 3LE.
\vskip2cm
\ec

\vskip1cm

\begin{abstract}
\noindent
We first consider local cosmic strings 
in dilaton-axion gravity and show that 
they are singular solutions.
Then we take a supermassive Higgs limit
and present expressions for the fields at
far distances from the core by 
applying a Pecci-Quinn and a duality transformation
to the dilatonic Melvin's magnetic universe.
\end{abstract}


\end{titlepage}


\section{Introduction}

Topological defects of various kinds are predicted
to have occurred during phase transitions in the
early universe. Depending on the nature of the phase
transition and the configuration of the fields involved
in spontaneous symmetry-breaking, topological defects can
be three, two or one dimensional, such as textures,
domain walls or strings, and also pointlike such
as monopoles. Their cosmological implications
have been quite well explored
in the context of general relativity
(for a review see \cite{Shellard}).
In particular the cosmic strings which have attracted the most
attention are those
produced at GUT scales. Such strings generate
large angle cosmic microwave background anisotropies
of roughly the observed order of magnitude \cite{0Hindmarsh}.
Their gravitational effects are weak and their
static spacetimes are non-singular \cite{2Linet,3Linet}.

On the other hand, at sufficiently high energy scales 
the Einstein's theory of general
relativity which is extremely successful
at describing the dynamics of our
solar system, and indeed the
observable universe, probably does not describe
gravity accurately \cite{RWILL}
as first postulated by Dirac
\cite{3Dirac}.

Partially motivated
by the Dirac's idea and the possible
existence of extra dimensions of   
the spacetime proposed by
Kaluza and Klein \cite{3Kaluza}, 
Jordan, Brans and Dicke (JBD) \cite{0Jordan}
proposed a theory
of gravity whose purpose    
was to incorporate
Mach's Principle. In this way
the variations on the inertial mass
of a body caused by the surrounding
Universe, assumed in that Principle,
could be justified from the variations of
the gravitational constant. Even so
the Principle is not consistently
justified, as it does  
not explain gravity in vacuum.

The implications of
the JBD action on cosmic string models
have been well explored \cite{3GOrtiz, 3DVilenkin}. 
In particular the case of the low-energy string action
was explored in \cite{1GregorySantos} which
showed that in the presence of a dilaton
the gravitational field of
cosmic strings is surprisingly close
to that of an Einstein cosmic string on
cosmological distance scales.
For a massless dilaton   
it was shown that there are
long range dilatonic effects on
the very large scale
and the spacetime is not
asymptotically locally flat
in the Einstein and string frames.
On scales of cosmological
interest, however, the metric is conical
in the Einstein frame, and
conformally conical in
the string frame.
Meanwhile for a massive
dilaton and apart from
distances of order the
Compton wavelength of the
dilaton, the long-range
structure of the string is as
for Einstein gravity, and so
the metric asymptotes a  
conical metric, in both    
string and Einstein frames. 
The main exception to these results
is for the special value where      
the coupling of the dilaton to the
vortex fields is canonical,
i.e. the vortex fields couple to the metric
in the string frame,
for which there are no
long range effects (other than
the deficit angle) and the
dilaton merely shifts the
value of the dilaton between
the core and infinity by
a constant. For a critical 
Higgs coupling this
constant vanishes and
there is no effect at all
on the dilaton. The dilatonic  
string is then the same as the
Einstein one.
 
Meanwhile, low energy string gravity also
generically contains another field - the axion -
which could in principle also couple to the cosmic string. 
Of course, in the cosmic super string solution
of \rm Dabholkar et al. \cite{3aGibbons} the axion
couples to the string.
However here we wish to consider the simple question
of whether and how, at the low energy level,
the presence of the axion effects the field of
a local vortex solution to some field theory,
using the $U(1)$ abelian Higgs theory as a model.

Therefore the layout of this paper is as follows:
Section 2 shows the singular behaviour
of the axionic-dilatonic local strings by
examining the gravi-axion-dilaton field.
Section 3 presents a far field solution 
by taking a supermassive
Higgs limit. Finally we conclude.

\section{Singular
axionic-dilatonic strings}

We are interested in the behaviour of
a static self gravitating
string whose metric is given by 
\be\label{1metricaestcilin}
ds^2 = e^\gamma\left[dt^2
- dR^2 -dz^2\right]-\alpha^2
e^{-\gamma} d\varphi^2\,\,.
\ee
when the gravitational interactions
take a form typical of low   
energy string theory
\cite{0Fradkin}.

We take an empirical
approach to cosmic strings in this
background theory, not concerning ourselves
with the origin of the fields that
form the vortex, but inputting ``$by$ $hand$''
the abelian-Higgs lagrangian given by
\be\label{12.1}
{\cal L}[\Phi ,A_c] = D_{a}\Phi ^{\dagger}D^{a}\Phi -
\frac{1}{4}{\tilde F}_{ab}{\tilde F}^{ab} -
\frac{\chi}{4}
(\Phi ^{\dagger} \Phi - \eta ^2)^2
\ee
where $D_c = \nabla _{c} + ieA_{c}$
is the usual gauge covariant derivative,
${\tilde F}_{ab}$ is the field strength
of the gauge field  $A_{c}$, $\eta$ is the symmetry
breaking scale and $\chi$ is a parameter related with
the energy density of the string, $\mu$, by
\be
\mu \sim \frac{1}{\sqrt{\chi}\, \eta}\,\,.
\ee 
To take account of the (unknown) coupling
of the cosmic string to the
dilaton , $\phi$, we consider a reasonably general
form for the interaction of the vortex
with the dilaton assuming that the
Abelian-Higgs lagrangian couples to the
dilaton via an arbitrary coupling,
$e^{2a\phi}\,{\cal L}$, in the string frame
as in the action:
\be\label{3stringaction}
\hat S = \int d^4 x
\sqrt{-\hat g} \left [ e^{-2\phi}
\left (-\hat R
-4({\hat \nabla} \phi)^2 
- {\hat V}(\phi)   
+\frac{1}{12}{\hat H}_{\mu\nu\lambda}^2 \right )
+ e^{2a\phi} {\hat {\cal L}}
\right ]
\ee
where ${\hat V}(\phi)$ is the dilaton potential
and the axion field is described by   
an antisymmetric tensor
$B_{\nu\lambda}$
with antisymmetric field strength
${\hat H_{\mu\nu\lambda}}$ given by:
${\hat H_{\mu\nu\lambda}}
= \partial_{[\mu}B_{\nu\lambda]}$.

We now write
instead the action
in terms of the ``$Einstein$'' metric,
defined as
\be
g_{ab} = e^{-2\phi} {\hat g}_{ab}
\ee
in which the gravitational
part of the action appears in the
more familiar Einstein form:
\be\label{3einsteinaction}
S = \int  d^4 x \sqrt{-g} \left [
- R + 2 (\nabla\phi)^2 - V(\phi)
+ \frac{1}{12} e^{-4\phi}
H_{\mu\nu\lambda}^2
+ 2 \epsilon e^{2(a+2)\phi} {\cal L}
\{X,P,e^{2\phi}g \} \right ]
\ee
and write the
``$Einstein$'' equations to get
\be\label{3Einsteinequation}
G_{ab} = \epsilon e^{2(a+2)\phi} T_{ab}
+ S_{ab} + N_{ab}
\ee
with $T_{ab}$, $S_{ab}$ and $N_{ab}$
the energy-momentum tensors for the
string, the dilaton and the axion
given respectivelly by:
\be\label{3.4}
T_{ab}=2 \frac{\delta
{\cal L}[X,P,e^{2\phi}g]}{\delta g^{ab}}
= 2e^{-2\phi} [ \nabla_aX\nabla_bX
+ X^2 P_aP_b] - 2\beta
e^{-4\phi} F_{ac}F_b^c - {\cal L} g_{ab}\,\,\,\,,
\ee
\be\label{33.6}
S_{ab} = 2 \nabla_a\phi\nabla_b\phi
+\frac{1}{2} V(\phi)g_{ab} - (\nabla \phi)^2
g_{ab}
\ee
and
\be\label{3axionenergymomentum}
N_{ab} = \frac{1}{12} e^{-4\phi}
[3H_{a{\lambda }k}H_{b}^{\,\,\,\,{\lambda} k}
-\frac{1}{2} g_{ab}H^2].
\ee
Considering the axionic antisymmetric
tensor, $B_{\nu\lambda}$,
as an independent variable, the
equation of motion for the axion is:
\be\label{3axionequation}
\nabla_{\mu}[e^{-4\phi}H^{\mu\nu\lambda}] = 0
\ee
which can be simplified into a wave
equation of motion for a massless
scalar field, $h(t,r,z,\varphi)$, 
evolving coupled to the dilaton
\cite{0CLWands}
\be\label{3axionwaveequation}
\Box h + 4 \nabla^{\mu}\phi\nabla_{\mu}h = 0
\ee
by writing
\be\label{3definitionofh}  
e^{-4\phi}H^{\mu\nu\lambda} =
\epsilon^{{\mu\nu\lambda}k}h_{,k}  
\ee
with $\epsilon^{{\mu\nu\lambda}k}$
the antisymmetric Levi Civita tensor given by
\be\label{3epsiloncontracted}
\epsilon_{abcd}\epsilon^{abcf} = -3! \delta_{d}^f.
\ee
In terms of $h$, the energy-momentum
tensor for the axion,
(\ref{3axionenergymomentum}),
then becomes:
\be\label{3energymomentumtensorforh}
N_{ab} = \frac{1}{2} e^{4\phi} [h_{,a} h_{,b}
- \frac{1}{2} g_{ab}\,\, h_{,\alpha} h_{,\beta}\,\, g^{\alpha\beta}].
\ee
To include the self-gravity of the string
we require a metric which   
exhibits the symmetries of the sources,
namely the boost invariance for
the matter and dilatonic sectors
($T^0_0 = T^z_z$, $S^0_0 =S^z_z$).
Therefore we try first the
static cylindrically
symmetric metric given in
(\ref{1metricaestcilin}). 
Consistency with the assumed symmetry
then requires
$N_{00} = -N_{zz}$, i.e.,
$h_{,t}=h_{,z}=0$ and also
$N_{R\varphi}=0$, i.e.,
$h_{,R}=0$ or $h_{,\varphi}=0$.
Therefore there are two
possible $Ans\ddot atze$ for the axion:
either
$h(R)$ or $h(\varphi)$.
Consistency of the constraint  
equation (given later by
(\ref{3eqnsgeometrydilatonh(R)c}))
also requires $N_R^R$ to be a
function of $R$ and therefore
for the ansatz $h(\varphi)$
we require $h_{,\varphi}=h$,
i.e., $h(\varphi)=
h \varphi$, with
$h$ a constant.

{\bf (i) $h(R)$}

Let us first assume $h(R)$.   
Integrating the equation of  
motion for the axion
(\ref{3axionwaveequation}) one gets:
\be\label{3equationforh(R)}
\alpha h^\prime = h_0 e^{-4\phi}
\ee
where $h_0$ is an integrating
constant, with
the energy-momentum tensor for the axion given
by:
${\hat N}_0^0 = {\hat N}_z^z
= {\hat N}_{\varphi}^{\varphi}
= - {\hat N}_R^R =
\frac{1}{4} e^{-\gamma+4\phi} h^{\prime2}$
where the prime means now the
derivative with respect to $R$.

The equations of motion for
the geometry
and for the
dilaton are respectively:
\bsq
\bq
&&\alpha^{\prime\prime}   
= -\alpha e^\gamma V(\phi)
-\epsilon\alpha e^\gamma
({\cal E} - {\cal P}_R)
\label{3eqnsgeometrydilatonh(R)a}\\
&&(\alpha\gamma^\prime)^\prime
= -\alpha e^\gamma V(\phi)  +
\epsilon\alpha
e^\gamma ({\cal P}_R + {\cal P}_\varphi)
\label{3eqnsgeometrydilatonh(R)b}\\
&&\alpha^\prime\gamma^\prime
= -\frac{1}{2} \alpha e^\gamma V(\phi)
+ \frac{\alpha \gamma^{\prime 2}}{4}
+ \alpha \phi^{\prime2} +
\epsilon\alpha e^\gamma {\cal P}_R
+\frac{h_0^2}{4\alpha} e^{-4\phi}
\label{3eqnsgeometrydilatonh(R)c}\\
&&(\alpha \phi^\prime)^\prime
= \frac{\alpha e^\gamma}{4}
\frac{\partial V}{\partial\phi}
+ \epsilon (a+1) \alpha
e^\gamma{\cal E} -
\frac{1}{2}\epsilon \alpha e^\gamma
({\cal P}_R + {\cal P}_\varphi)
+\frac{h_0^2}{2\alpha} e^{-4\phi}
\label{3eqnsgeometrydilatonh(R)}
\eq
\esq
while the equations of motion for the
vortex sector are given by
\bsq
\bq
&&\frac{1}{\alpha}\,\left(\alpha \,
X^\prime \right)^\prime
= \frac{X\,P^2}{\alpha^2}\,e^{2\gamma}
+ \frac{X}{2}\,(X^2-1)\,e^{\gamma}
\label{12.10d}\\
&&\alpha\,\left(\frac{P^\prime}{\alpha}
\right)^\prime
= - \gamma^\prime \, P^\prime +
\frac{X^2\,P}{\beta} \,e^{\gamma}
\label{12.10e}
\eq
\esq
-the same as for the dilatonic
string, as the axion does not
couple directly to matter. 
For further reference,
the Bianchi identity now gives:
\bq
&&\epsilon(\alpha e^\gamma {\cal P}_R)^\prime
-\epsilon\alpha^\prime e^\gamma {\cal P}_\varphi
-\frac{\epsilon}{2}\alpha\gamma^\prime e^\gamma
[ {\cal P}_R - {\cal P}_\varphi - 2{\cal E} ]
\nonumber\\
&&+\alpha^\prime \phi^{\prime 2}
+ (\alpha \phi^{\prime 2})^\prime
-\frac{\alpha}{2} e^\gamma \phi^\prime
\frac{\partial V}{\partial\phi}
+\alpha e^{4\phi} h^\prime
[h^\prime (\frac{\alpha^\prime}{2\alpha}
+\phi^\prime) +\frac{h^{\prime\prime}}{2}]
= 0
\label{3Bianchih(R)}
\eq 
Let us start to look for the string's
gravitational effects at  
very far distances from its core,
where for non-singular  
local strings we are in vacuum.

Consider first the massless dilaton
case, $V(\phi)\equiv0$ for which
the equations of motion for the
geometry and the dilaton become:
\bsq
\bq
&&\alpha^{\prime\prime} =0  \\
&&(\alpha\gamma^\prime)^\prime = 0 \\
&&\alpha^\prime\gamma^\prime
= \frac{\alpha\gamma^{\prime 2}}{4}
+ \alpha\phi^{\prime2}
+\frac{h_0^2}{4\alpha} e^{-4\phi}  
\label{3farequsgeomdilatonh(R)} \\
&&(\alpha \phi^\prime)^\prime =
\frac{h_0^2}{2\alpha} e^{-4\phi}.
\label{3fareqsgeomdilatonh(R)}
\eq
\esq
These equations are
only consistent for
the trivial case where
the axion is
``$\it switched$ $\it off$''
and so the results in 
\cite{1GregorySantos}
for the dilatonic string still hold.
Indeed one gets:
\bsq
\bq
&&\alpha  =  d R + b
\label{alpham=0}\\
&&\gamma  =  C_3 + \frac{C_2}{d}
Ln\left[d R + b\right]
\label{gammam=0}\\
&&\phi = Ln\left[e^{-8\phi_0}
\frac{8C}{\left(Cosh(A)\right)^2}
\right]^{\small -\frac{1}{4}}
\label{3solutionsh(R)}
\eq
\esq
with
\be\label{3defenitionofA}
A = \frac{\sqrt{8C}}{d}
Ln\left[\frac{d R + b}{C_1}\right]
\ee
where $C$, $C_1$, $C_2$, $C_3$,
$d$ and $b$ are integrating constants.

From the dilaton equation
(\ref{3fareqsgeomdilatonh(R)})
it comes that
$C$ is defined
such that:
\be\label{3definitonofC}
(\alpha \phi^\prime)^2
= -\frac{h_0^2}{4} e^{-4\phi}
+ 2C \geq0
\ee
and therefore $C > 0$.
But now we find a contradiction:-
integrating (\ref{3eqnsgeometrydilatonh(R)b})
from the core and to lowest order in
$\epsilon$ one gets $\alpha \gamma^\prime=0$,
i.e., $C_2=0$. Using the constraint
(\ref{3farequsgeomdilatonh(R)}),   
$dC_2 = \frac{1}{4} C_2^2 + 2C$,
this gives $C=0$ which contradicts 
the fact that $C > 0$.  
This means $h^\prime=\phi^\prime=0$.
Therefore the only
vacuum solution is the
trivial one
where the axionic sector
is ``$\it switched$ $\it off$''
and the dilaton is constant.
This result still holds
in the string frame and is
in agreement with those
presented in reference
\cite{3aGibbons} for massless
dilatons in a spacetime
of four dimensions.

Consider now massive dilatons
($V(\phi)=2M^2 \phi^2$, with $M$
the dilaton mass).
Apart from an intermediate
annular region (bounded by
the Compton wavelength of
the dilaton), the long-range
structure of the fields is as for
the Einstein case because for distances
from the core greater than the   
Compton wavelength of the dilaton the
dilaton field is essentially fixed.
Hence the equations of motion for
the geometry and the dilaton become:
\bsq
\bq
&&\alpha^{\prime\prime}
=-\alpha e^\gamma 2 M^2 \phi^2  \\
&&(\alpha\gamma^\prime)^\prime =
-\alpha e^\gamma 2 M^2 \phi^2 \\
&&\alpha^\prime\gamma^\prime
= -\alpha e^\gamma M^2 \phi^2
+\frac{\alpha\gamma^{\prime 2}}{4}
+\frac{h_0^2}{4\alpha} e^{-4\phi}
\label{3farequsgeomdilatonh(R)M} \\
&&0=\alpha e^\gamma M^2 \phi
+\frac{h_0^2}{2\alpha} e^{-4\phi}.
\label{3fareqsgeomdilatonh(R)M}
\eq
\esq
whose solution is:
\bsq
\bq
&& \alpha-\alpha_0 = C_4 R\\
&& \gamma-\gamma_0 = -2 \ln[\alpha_0 + C_4 R] \\
&& h - h_1 = -2 \phi_1 \frac{C_5}{C_4} \ln[\alpha_0 + C_4 R]
\label{axionhR}\\
&& \phi = \phi_1
\eq
\esq
with
\be
-\frac{h_0^2 e^{-4\phi_1}}{2\phi_1}=C_5=\alpha^2e^\gamma M^2
\ee

which from (\ref{3farequsgeomdilatonh(R)M}) gives
\be
\phi_1 = -\frac{1}{4}-\frac{1}{4} \sqrt{1+48\frac{C_4^2}{C_5}}
\ee
where $\alpha_0$, $C_4$, $\gamma_0$, $h_1$ are integrating
constants and $C_5$ is a positive constant.

The asymptotic solution for the axionic-dilatonic
cosmic string is then, to lowest order:
\be
ds^2=e^{\gamma_0}\left[\alpha_0+C_4 R\right]^{\small -2}
\left[ dt^2 - dR^2 -dz^2\right]
- e^{-\gamma_0}\left[\alpha_0+C_4 R\right]^{\small 4} d\varphi^2
\ee
with $\phi=\phi_1$ and the axion having very strong
asymptotic effects as given in (\ref{axionhR}).

{\bf (ii) $h(\varphi)=h\varphi$}

Let us now consider the
other ansatz for the axion,      
$h(\varphi)=h\varphi$.
The axion is no more a
dynamical field and so its
equation of motion
gives an identity.
The elements of the
energy-momentum
tensor are:
${\hat N}_0^0 = {\hat N}_z^z
= {\hat N}_R^R
= - {\hat N}_{\varphi}^{\varphi}
=\frac{1}{4\alpha^2} e^{4\phi}
h^2 e^{\gamma}$ and the
the axionic sector
in the initial action
(\ref{3einsteinaction})
is given by
$\frac{1}{12} e^{-4\phi}
H_{\mu\nu\lambda}^2=
\frac{1}{2} e^{4\phi+\gamma}
\frac{h^2}{\alpha^2}$.

The equations of motion for
the geometry and the dilaton are now
respectively:
\bsq
\bq
&&\alpha^{\prime\prime} =
-\alpha e^\gamma V(\phi)
-\epsilon\alpha e^\gamma ({\cal E} - {\cal P}_R)
- \frac{h^2}{2\alpha}
e^{2\gamma+4\phi} \label{31eqnsgeometrydilatonh(theta)} \\
&&(\alpha\gamma^\prime)^\prime
= -\alpha e^\gamma V(\phi)
+ \epsilon\alpha e^\gamma ({\cal P}_R
+ {\cal P}_\varphi)
\label{32eqnsgeometrydilatonh(theta)} \\
&&\alpha^\prime\gamma^\prime =
-\frac{1}{2} \alpha e^\gamma V(\phi) +
\frac{\alpha \gamma^{\prime 2}}{4}
+ \alpha \phi^{\prime2} +
\epsilon\alpha e^\gamma {\cal P}_R
-\frac{h^2}{4\alpha} e^{2\gamma+4\phi} 
\label{33eqnsgeometrydilatonh(theta)} \\
&&(\alpha \phi^\prime)^\prime =
\frac{\alpha e^\gamma}{4}
\frac{\partial V}{\partial\phi}
+ \epsilon (a+1) \alpha e^\gamma {\cal E} -
\frac{1}{2}\epsilon \alpha e^\gamma ({\cal P}_R + {\cal P}_\varphi)
+\frac{h^2}{2\alpha} e^{2\gamma+4\phi}
\label{3eqnsgeometrydilatonh(theta)}
\eq
\esq
and again the equations of motion for
the vortex fields are the same as
for the dilatonic string.
The Bianchi identity is now:
\bq
&&\epsilon(\alpha e^\gamma {\cal P}_R)^\prime
-\epsilon\alpha^\prime e^\gamma {\cal P}_\varphi
-\frac{\epsilon}{2}\alpha\gamma^\prime e^\gamma
[ {\cal P}_R - {\cal P}_\varphi - 2{\cal E} ]
\nonumber\\
&&+\alpha^\prime\phi^{\prime2}
+ (\alpha \phi^{\prime2})^\prime
-\frac{\alpha}{2} e^{\gamma} {\phi}^\prime
\frac{\partial V}{\partial\phi}
- \frac{h^2}{\alpha} e^{4\phi+2\gamma}
\phi^\prime = 0.  
\label{3Bianchih(theta)}
\eq
Again, consider first the massless case,
$V(\phi)\equiv0$.
Proceeding as for the previous ansatz we
write the
equations of motion for the geometry
and the dilaton:
\bsq
\bq
&&\alpha^\prime = - \alpha\phi^\prime
+ q  \label{31fargeomdilatonh(theta)} \\
&&\alpha\gamma^\prime =p
\label{32fargeomdilatonh(theta)} \\
&&\alpha\frac{\alpha^{\prime\prime}}{2}
+ \alpha^\prime (- 2 q - p)
+ (q^2 + \frac{1}{4} p^2) +
\alpha^{\prime 2} =0 \label{33fargeomdilatonh(theta)} \\  
&&(\alpha \phi^\prime)^\prime =
\frac{1}{2\alpha} h^2 e^{2\gamma+4\phi}
\label{3fargeomdilatonh(theta)}
\eq
\esq
at far distances from the core
where it is vacuum, with
the Bianchi identity giving
\be\label{3farBianchih(theta)}
\alpha^\prime\phi^{\prime2}
+ (\alpha \phi^{\prime2})^\prime
- \frac{h^2}{\alpha}
e^{4\phi+2\gamma}\phi^\prime = 0
\ee
where $p$ and $q$ are integrating constants.
In order to get the asymptotic solutions   
to those equations we write the constraint
(\ref{33fargeomdilatonh(theta)})
as the autonomous d.s.:
\bsq
\bq 
&&u^\prime=-3u^2+2(2q+p)ut
-2(q^2+\frac{p^2}{4})t^2 \\
&&t^\prime = -ut
\eq
\esq
where $t =
\frac{1}{\alpha}$ and
$u=\frac{\alpha^\prime}{\alpha}$.
Writing $x = \frac{u}{t} =\alpha^\prime$
one gets:  
\be
dx=\left(3x+\frac{2}{x}(q^2+\frac{p^2}{4})
-2(2q+p) \right)\,\frac{dt}{t}
\ee
which integrated gives
\be\label{3t(x)}  
t(x) = C \left(Q(x)\right)^{\small \frac{1}{6}}
Exp\left[-\frac{\sqrt2 (p+2q)}{3\sqrt{p^2-8pq+4q^2}}
\arctan[\frac{\sqrt2(p+2q-3x)}{\sqrt{p^2-8pq+4q^2}}]
\right]
\ee
with $C$ a positive integrating constant and
$Q(x) = p^2+4q^2-4px-8qx+6x^2$,
a non-invertible expression for
$\alpha(R)$
whose asymptotic regimes are then
analysed.

When $|x| \rightarrow \infty$ the result
from expression (\ref{3t(x)})
is that $t\rightarrow \infty$ where
\be\label{3xinfty}
t(x)=\frac{4}{C_3^4}|x|^{\small \frac{1}{3}}
\ee
Therefore the solution is:
\bsq
\bq
&&\alpha  = \alpha_0
+C_3R^{\small \frac{1}{4}}
\label{3solutionbranche1a}\\  
&&\gamma  =  \gamma_0   
+p\left(
\frac{4\alpha_0^2}{C_3^3}R^{\small \frac{1}{4}}
-\frac{2\alpha_0}{C_3^2} \sqrt{R}
+ \frac{4}{3C_3}R^{\small \frac{3}{4}}
-\frac{4\alpha_0^3}{C_3^4}Ln\left[
\alpha_0+C_3R^{\small \frac{1}{4}}
\right]
\right)
\label{3solutionbranche1b} \\
&&\phi = \phi_0
+q\frac{4\alpha_0^2}{C_3^3}R^{\small \frac{1}{4}}
-q\frac{2\alpha_0}{C_3^2} \sqrt{R}
+q \frac{4}{3C_3}R^{\small \frac{3}{4}}
-(1+q\frac{4\alpha_0^3}{C_3^4})Ln\left[
\alpha_0+C_3R^{\small \frac{1}{4}}
\right]
\label{3solutionbranche1c}
\eq
\esq
with $C_3$ a positive constant
and $\alpha_0$, $\gamma_0$ and
$\phi_0$ being integrating constants.
But now an inconsistency comes-
consistency of equation
(\ref{3fargeomdilatonh(theta)})
requires $C_3=h=0$, i.e., the
axionic sector for these solutions
is ``$\it switched$ $\it off$''.

When $x \rightarrow x_{\pm}$,
where $Q(x_{\pm})=0$, therefore
from expression (\ref{3t(x)}) $t \rightarrow 0$.
As a result the solution is:
\bsq   
\bq
&&\alpha =  \alpha_0 + x_{\pm} R \\
&&\gamma  = \gamma_0 + \frac{p}{x_{\pm}}
Ln\left[\alpha_0+x_{\pm} R\right] \\
&&\phi  = \phi_0 + (\frac{q}{x_{\pm}}-1)
Ln\left[\alpha_0 + x_{\pm} R\right]
\label{3solutionbranche2c}
\eq
\esq
which from (\ref{3fargeomdilatonh(theta)}),
as in previous asymptotic regimes, 
gives $h=0$, i.e.,
the axionic sector for these solutions
it is as well ``$\it switched$ $\it off$''.

Finally when $x \rightarrow 0$ then,
from expression (\ref{3t(x)}) $t \rightarrow t(0)$
and the solution is:
\bsq
\bq
&&\alpha =  \alpha_0 \\ 
&&\gamma  = \gamma_0 + \frac{p}{\alpha_0}R \\
&&\phi  = \phi_0 + \frac{q}{\alpha_0} R
\label{3solutionbranche3c}
\eq
\esq
which, again, from (\ref{3fargeomdilatonh(theta)})
gives $h=0$.
In conclusion the only vacuum solution
is the trivial one where the axion
is ``$\it switched$ $\it off$''.
The same result
is obtained in the string frame.  
Moreover no solution
(in the Einstein frame)
is regular at the
origin where ${\hat N}_0^0
= \frac{1}{4\alpha^2}
e^{4\phi+\gamma} h^2
\sim \frac{1}{R}$,
i.e., $e^{4\phi}
\sim R$
because
from the dilaton equation
(\ref{3eqnsgeometrydilatonh(theta)})
one obtains $\phi \sim R$
whilst from the constraint   
(\ref{33eqnsgeometrydilatonh(theta)})
one gets $\phi \sim \sqrt{R} $. 
Therefore solutions of the equations
of motion are non-string type solutions.

It is worth examining the
previous results.
First we note that for the
ansatz $h(\varphi)=h\varphi$ the
elements of the energy-momentum
tensor for the axion resemble those
for a global string given by
\be
{\hat T}^i_i = \frac{1}{R^2}\,
\left[1,\,\,\,\,1,\,\,\,\,1,\,\,\,\,-1\right]
\label{1tensorglobal}
\ee
and for which the energy per unit length
is infinite, i.e., the string field has 
long-range gravitational interactions.
This fact when combined with the
absence of non-trivial vacuum solutions
suggest that fields for the axion-dilatonic string
also have a long-range behaviour.
Meanwhile taking a supermassive Higgs limit
at far distances from the core of a string
in Einstein gravity the boson vectorial mass vanishes
and one is left with the
Einstein-Maxwell equations which can be solved
exactly. The Higgs boson becoming supermassive
decouples from the eletromagnetic field and
the metric solution is the Melvin magnetic Universe
given by \cite{3Melvin}
\be\label{mmelvin}
ds^2 = (1+\frac{1}{4} B^2 R^2)^2 \left [
dt^2 - dR^2 - dz^2 \right ]
- \frac{R^2}{(1+\frac{1}{4} B^2 R^2)^2}
d\varphi^2
\ee
with $B$ the strength of the magnetic field
along the axis of symmetry.
This suggests that in the presence of the axion and the
dilaton one shall look for a supermassive 
Higgs limit and one shall obtain an exact
metric solution - the dyonic Melvin universe
which will be derived in the next section.

We also note that the asymptotic limit of
the solution (\ref{mmelvin})
is of the form:
\be
ds^2 = \rho^4\,\left[
dt^2-dz^2-d\rho^2
\right]-\frac{D^2}{\rho^2}\,d\varphi^2
\ee
with $D$ a positive constant which is 
classified as a ``$global\,\,\,\,string$''
in the sense of \rm Comtet and Gibbons 
\cite{3CGibbons} - a static
solution of the vacuum Einstein equations which
is invariant under boosts along the direction
of the string and which is invariant under
translations and reflections in a plane orthogonal
to the translationns together with rotations 
about an axis of symmetry and reflections 
in planes through the axis, i.e., 
having ``$whole\,\,\,\,cylinder-symmetry$'' and
being non asymptotically locally flat. Of course,
the metric of a global vortex (generated by some
field theory) does not have this form \cite{14Gregory}
but the form
\be
ds^2=b_0(r-r_0)^2\left[dt^2-\cosh^2\sqrt{b_0}\,\,t\,\,dz^2
\right]-dr^2-C_0^2 d\varphi^2
\ee 
with $b_0$ and $C_0$ two constants ($b_0>0$)
which is clearly non-static (if we force the metric
to be static it becomes singular 
\cite{Kaplan}) and has an event horizon at $r=r_0$.
Also note that in the presence of a dilaton the asymptotic
form of a global vortex is also non-static
as was shown in \cite{1OGregory}.

\section{\bf Global string solutions: dyonic universes}

{\bf Dilatonic Melvin Universe}

\bigskip

The dilatonic Melvin magnetic universe
is an exact solution of the
Dilatonic-Einstein-Maxwell theory
described by the action:
\be\label{3bosoniceinsteinaction}
S = \int  d^4 x \sqrt{-g} \left [
-R + 2 (\nabla\phi)^2 - e^{-2a\phi} F^2\right ]
\ee
where $F$ is the field strength of any
$U(1)$ gauge field, $A$.

In the presence of a magnetic
field $B$ the equations of motion
given by \cite{3Dowker}:
\bsq
\bq
&&\nabla_{\mu}\left[e^{-2a\phi}
F^{\mu\nu}\right] = 0 \label{3nome1A}\\
&&\Box\phi+\frac{1}{2} a e^{-2a\phi} F^2 = 0
\label{3nome2A}\\
&&R_{\mu\nu} = 2 \nabla_{\mu} \phi
\nabla_{\nu} \phi + 2 e^{-2a\phi}
F_{\mu\rho} F_{\nu}^{\rho}
- \frac{1}{2} g_{\mu\nu} e^{-2a\phi} F^2.
\label{3nome3A}
\eq
\esq
admit an exact solution, the dilatonic
Melvin magnetic universe, given by
\cite{3Dowker}
\bsq
\bq
&&ds^2 = \left(1+\frac{a^2+1}{4} B^2 R^2\right)^{\small \frac{2}{1+a^2}}
\left [dt^2 - dR^2 - dz^2 \right ]
-\left(1+\frac{a^2+1}{4} B^2 R^2\right)^{\small -\frac{2}{1+a^2}} R^2
d\varphi^2
\nonumber\\
&&\,\,\,\,\,\,\,\,
\label{3nome1H}\\
&&e^{-2a \phi} = \left(1+\frac{a^2+1}{4}
B^2 R^2\right)^{\small \frac{2a^2}{1+a^2}} \label{3nome2H}\\
&&A_\varphi=-\frac{2}{(1+a^2)B(1+\frac{a^2+1}{4} B^2 R^2)}\label{3nome3H}\\
&&B^z=-\frac{B}{\left(
1+\frac{a^2+1}{4}B^2R^2\right)^{\small 2\frac{(2+a^2)}{1+a^2}}}
\eq
\esq
The equations of motion
are duality invariant as they admit an $SL(2,R)$
electric-magnetic duality defined by:
\bsq
\bq
&&\phi^\prime = - \phi \label{3nome1B}\\
&& F^\prime_{\mu\nu} = \frac{1}{2} e^{-2a\phi} \epsilon_{\mu\nu\rho\sigma}
F^{\rho\sigma} \label{3nome2B}
\eq
\esq
which can be used to generate
an electromagnetic Melvin universe
and that will be generalised in
the presence
of an axion in the next section. 

\bigskip
   
{\bf Dyonic Melvin Universe}   

\bigskip

In the presence of an axion,
the Melvin magnetic universe
can be generated after
applying a Peccei-Quinn and a duality transformation
to the dilatonic Melvin magnetic universe.
Our starting point comes from the bosonic part of the four-dimensional
effective
action of low energy string theory that includes the terms \cite{3Shapere} :
\be\label{3baxioneinsteinaction}  
S = \int  d^4 x \sqrt{-g} \left [
-R + 2 (\nabla \phi)^2 + \frac{1}{12} e^{-4\phi} H_{\mu\nu\lambda}^2
- e^{-2\phi} F^2
\right ]
\ee
where $H_{\mu\nu\lambda}$ describes the axion, with now
\be
H = \partial B + \frac{1}{4} A \wedge F.
\ee
If we write the axion as in
(\ref{3definitionofh}) the equations
of motion can equally be derived
from the action \cite{3Shapere} :
\be
S = \int dx^4 \sqrt{-g} \left[
-R+2(\nabla\phi)^2
+ \frac{1}{2} e^{4\phi} (\nabla h)^2
- e^{-2\phi}F^2
-h\,F\,{\tilde F}
\right]
\ee 
with ${\tilde F}$ the dual of $F$.

It is also convenient to combine the axion, $h$, and the
dilaton, $\phi$, in a complex scalar field, $\lambda
= \lambda_1 + i \lambda_2 = h + ie^{-2\phi}$
called the ``$axidilaton$''.
In terms of $\lambda$ the action simplifies to \cite{3Shapere} :
\be\label{3actionlambda}
S = \int  d^4 x \sqrt{-g} \left [
-R+\frac{|\nabla \lambda|^2}{2\lambda_2^2}
+\frac{i}{4}(\lambda F_{+}^2-{\bar \lambda} F_{-}^2)
\right ]
\ee
where $F_{\pm}=F\pm i{\tilde F}$.
The kinetic term for $\lambda$ in
(\ref{3actionlambda}) is invariant
under the Peccei-Quinn shift of
the axion, $h \rightarrow h + f$
and under the duality transformation
$\lambda \rightarrow -\frac{1}{\lambda}$.
The latter, for $h=0$, reduces
to the $\phi \rightarrow -\phi$
transformation of (\ref{3nome1B}).

The action as a whole is not invariant under this
duality, because the Maxwell terms
change sign, therefore the equations
of motion given by \cite{3Shapere}:
\bsq
\bq
&&\frac{\nabla_{\mu} \partial^{\mu} \lambda}{\lambda_2^2}
+i\frac{\partial_{\mu} \lambda \partial^{\mu} \lambda}{\lambda_2^3}
-\frac{i}{2}F_{-}^2 = 0 \label{3nome1C}\\
&&\nabla_{\alpha}\left[\lambda F_{+}^{\alpha\beta}
-{\bar \lambda} F_{-}^{\alpha \beta} \right] = 0 \label{3nome2C}\\
&&R_{\mu\nu}=\frac{\partial_{\mu}{\bar \lambda} \partial_{\nu} \lambda
+ \partial_{\nu}{\bar \lambda} \partial_{\mu} \lambda}{4\lambda_2^2}
+ 2 \lambda_2 F_{\mu\rho} F_{\nu}\,^{\rho}
- \frac{1}{2} \lambda_2 \,g_{\mu\nu} F^2
\label{3nome3C}
\eq
\esq
are not invariant either. In fact the Einstein equation
(\ref{3nome3C}) is not invariant in general under
the duality transformation \cite{3Shapere}
\bsq
\bq
&&\phi^\prime = - \phi \label{3name1A}\\
&&F^\prime_{+} = - \lambda F_{+} \label{3name2A}\\
&&F^\prime_{-} = - {\bar \lambda} F_{-} \label{3name3A}
\eq
\esq
as the second and third terms are non-invariant \cite{3Shapere}
and transform into \cite{3Shapere}:
\be
2\lambda_2F_{\mu\rho} F_{\nu}\,^{\rho}
-\frac{1}{2} \lambda_2 g_{\mu\nu} F^2
-\frac{\lambda_1\lambda_2^2}{|\lambda|^2}\left(
2F_{\mu\rho} {\tilde F}_{\nu}\,^{\rho}
+ 2 F_{\nu\rho} {\tilde F}_{\mu}\,^{\rho}
- g_{\mu\nu} F {\tilde F}\,\,.
\right)
\ee
Only if the extra terms in this   
expression vanish will the duality
transformation map solutions into
solutions. A duality transformation 
of the offending terms above shows that  
they transform into themselves with
\be
2F_{\mu\rho} {\tilde F}_{\nu}\,^{\rho}
+ 2 F_{\nu\rho} {\tilde F}_{\mu}\,^{\rho}
- g_{\mu\nu} F {\tilde F}\rightarrow
\left(\lambda_1^2-\lambda_2^2
\right)\left(
2F_{\mu\rho} {\tilde F}_{\nu}\,^{\rho}
+ 2 F_{\nu\rho} {\tilde F}_{\mu}\,^{\rho}
- g_{\mu\nu} F {\tilde F}
\right)
\ee
and therefore the Einstein equation is invariant when
\cite{3Shapere}
\be\label{3nome1D}
2F_{\mu\rho} {\tilde F}_{\nu}\,^{\rho}
+ 2 F_{\nu\rho} {\tilde F}_{\mu}\,^{\rho}
- g_{\mu\nu} F {\tilde F} = 0
\ee
which implies that if this expression vanishes
for a particular solution, it will also vanish for its duality
transformation of that solution. Thus, beginning
with a solution one may build up a family of solutions.
Indeed the Melvin magnetic universe verifies the restriction
(\ref{3nome1D}) (the only non-vanishing field strength is
$F_{R\varphi}=\frac{RB}{(1+\frac{B^2 R^2}{2})^2}$)
and therefore one can generate a axionic-dilatonic Melvin electromagnetic
universe.
To do this we start from the dilatonic Melvin magnetic solution
(\ref{3nome1H})
- (\ref{3nome2H}) and obtain an axion from a shift $f$ on a Peccei-Quinn
transformation, $\lambda' = \lambda + f$, then dualise it,
$\lambda' = -\frac{1}{\lambda}$, and finally rescale, $\lambda' = \lambda (f^2+1)$
\cite{3Shapere}.
One gets \cite{3Shapere} :
\bsq
\bq
&&\lambda^\prime
= -\frac{1}{\lambda + f} \left[f^2 + 1\right]
\label{3name1B}\\
&&h^\prime
= -\frac{f}{f^2 + e^{-4\phi}} \left[f^2 + 1\right] \label{3name2B}\\
&&e^{-2\phi^\prime}
= \frac{e^{-2\phi}}{f^2 + e^{-4\phi}} \left[f^2 + 1\right]
\label{3name3B}\\
&&F^\prime
= \frac{1}{\sqrt{f^2 + 1}} \left[-f F+e^{-2\phi} {\tilde F}\right]
\label{3name4B}\\
&&A^\prime _\varphi = -\frac{f}{\sqrt{f^2+1}} A_\varphi
\eq
\esq
with $A^\prime_z$, $A^\prime_t$ and $A^\prime_R$ unknown.
For the dilatonic Melvin magnetic universe $F_{R\varphi}
=\frac{RB}{k^2}$ and therefore:
\bsq
\bq
&&F^\prime_{tz}
= \frac{e^{-2\phi}}{k^2 R} B_1 \label{3name1C}\\
&&F^\prime_{R\varphi}
= \frac{R}{k^2} B_2 \label{3name2C}
\eq
\esq
with
\bsq
\bq
&&B_1 = \frac{1}{\sqrt{f^2 + 1}} B \label{3name1D}\\
&&B_2 = -\frac{f}{\sqrt{f^2 + 1}} B \label{3name2D}
\eq
\esq
where $B_2$ is the magnetic field of the axionic sector
and $B^2 = B_1^2 + B_2^2$ with $k=1+\frac{B^2\,R^2}{2}$.

Using (\ref{3nome2H}) with $a=1$ to rewrite the dilaton before
the transformation one finally gets the exact dyonic Melvin electromagnetic
universe, with the axion and dilaton fields given by:
\bsq
\bq
&&h(B_1,B_2,R) = \frac{B_2 B_1}{B_1^2+B_2^2}
\frac{1}{\left(1+\frac{1}{4} R^4 \left(B_1^2+B_2^2 \right)B_1^2
+ R^2 B_1^2\right) } \label{3name1E1}\\
&&e^{-2\phi (B_1,B_2,R)} = \frac{B_1^2}{B_1^2+B_2^2}
\frac{1+\frac{1}{2} \left(B_1^2+B_2^2 \right) R^2}
{\left(1+\frac{1}{4} R^4 \left(B_1^2+B_2^2 \right)B_1^2+ R^2 B_1^2
\right) }
\label{3name1E2}
\eq
\esq
with the metric given by:
\be\label{3name1F}
ds^2 = k\,\left[dt^2 - dR^2 - dz^2 \right ]
-\frac{1}{k}\,R^2\,d\varphi^2.
\ee

As expected this solution is consistent with the
dilatonic Melvin magnetic universe, (\ref{3nome1H}) - (\ref{3nome2H}),
which corresponds to the case where
there is no axion ($f = 0$)
and so $B_1 = B$ ($B_2 = 0$)
with $e^{-2\phi^\prime} \equiv e^{2\phi} =\frac{1}{1+\frac{1}{2} B^2 R^2}$.

\bigskip

{\bf Asymptotic global string solutions}
   
\bigskip

At very far distances from the core
the field solutions are then 
obtained from (\ref{3name1E1})-(\ref{3name1E2})
by taking the limit
$R \rightarrow \infty$. Therefore the
axion and the dilaton are given by
\bsq
\bq
&&h(B_1,B_2,R) \rightarrow \frac{4 B_2}{R^4\,B^4 B_1}\\
&&\phi (B_1,B_2,R) \rightarrow \ln \left[
\frac{B\,R}{\sqrt{2}}\right]
\eq
\esq
with the metric for the axionic-dilatonic
global string given by
\be
ds^2= \left(\frac{BR}{\sqrt2}\right)^2 \,\left[
dt^2 - dR^2 - dz^2 \right ]
-\frac{2}{B^2}\,d\varphi^2.
\ee

\section{\bf Conclusions}

In the presence of an axion 
we showed that the gravitational
field of the string is quite different
from the Einstein and dilatonic ones \cite{1GregorySantos}.
In fact the long range effects of the dilaton and
of the axion exclude the existence
of local string solutions.
Instead we looked for far field
solutions and presented expressions
for the asymptotical fields which
approaches the dilatonic Melvin
magnetic universe in the
presence of an axion.
The axion is strongly damped to
zero while the dilaton has very strong
asymptotic effects similar to those
for the dilatonic string for massless
dilaton.

{\bf Acknowledgements}

The author would like to thank Dr. Ruth Gregory
for reading this manuscript, for useful discussions
and suggestions.

\end{document}